# New proximity potential for alpha decay for superheavy nuclei

H.C.Manjunatha[1]*, K.N.Sridhar[2]
[1]Department of Physics, Government College for Women, Kolar-563101 Karnataka, INDIA
[2]Department of Physics, Government First Grade College, Kolar-563101 Karnataka, INDIA
*Corresponding Author: manjunathhc@rediffmail.com

## Abstract

We have constructed new proximity function particularly for interaction between two superheavy nuclei based on the experimental alpha decay half-lives. The new proximity function is used to produce the alpha decay half-lives of superheavy nuclei whose experimental values are known. The new proximity function produces the alpha decay half-lives close to the experiments. Hence we can conclude that the new proximity function can be used to study the interaction between two superheavy nuclei.

**PACS numbers:** 24.10.-i; 25.70.Jj; 25.60.Pj; 25.70.-z, 23.70.

## I. Introduction

The accurate determination of the interaction potential between two nuclei is different task and researchers are involved in this direction. The Coulomb repulsion force is a long range and nuclear attractive force is a short range force. Coulomb part of interaction is well known. Whereas nuclear part is not clearly understood. From the literature survey, it is observed that there has been many efforts to calculate nuclear potentials [1-11]. Among such methods, proximity potential is found to be simple and accurate. Shi and Swiatecki [12] were the first to use the proximity potential. Proximity potential is composed of two parts. One depends on shape and geometry of two nuclei. The other is the universal function $\phi(s)$ only related to the distance of separation between two nuclei. The universal function is independent of shapes of two nuclei and geometry of nuclei. The idea of universal function is fundamental advantage of proximity potential. The exact value of nuclear potential depends on the barrier penetrability which intern depends on proximity function. The different proximity potentials used in the study of heavy ion fusion and decay process. Bass in 1977 [13] introduced nuclear part of interaction potential based on experimental data of fusion cross sections by adopting liquid drop model and geometric interpretations and is given by

$$V_N(r) = -\frac{R_1 R_2}{R_1 + R_2} \phi(s)$$

The universal function is given by

$$\phi(s) = [0.03 \exp(s/3.3) + 0.0061 \exp(s/0.65)]^{-1}$$

Ngo 1980 (Ng80) [14] evaluated nuclear potential using the following universal function expression (prox 1980).

$$\phi(s) = \begin{cases} -33 + 5.4(s-s_0)^2 & \text{for } s < s_0 \\ -33 \exp\left[-\frac{1}{5}(s-s_0)^2\right] & \text{for } s > s_0 \end{cases}$$

Reisdorf, proposed the following universal function [15] which is known as prox 1994 to evaluate the nuclear potential.

$$\phi(s) = [0.033\exp(s/3.5) + 0.007\exp(s/0.65)]^{-1}$$

To evaluate nuclear potential [16], Blocki et al., proposed the following expression

$$V_N(r) = 4\pi \bar{R} \gamma b \phi(s)$$

Where, $\phi(s)$ is the universal function [16], (prox 77)

$$\phi(s) = \begin{cases} -\frac{1}{2}(s-2.54)^2 - 0.0852(s-2.54)^3 & for \quad s < 1.2511 \\ -3.437\exp(-s/0.75) & for \quad s > 1.2511 \end{cases}$$

Blocki et al., evaluated the nuclear proximity potential using the following universal function [17] also known as GP 77.

$$\phi(s) = \begin{cases} -1.7817 + 0.927s + 0.0169s^2 - 0.0514s^3 & for \quad 0 < s < 1.9475 \\ -4.41\exp(-s/0.7176) & for \quad s > 1.9475 \end{cases}$$

The universal function $\phi(s)$ used by Blocki et al., [17] to evaluate the nuclear potential (prox 81)

$$\phi(s) = \begin{cases} -1.7817 + 0.927s + 0.143s^2 - 0.09s^3 & for \quad s < 0 \\ -1.7817 + 0.927s + 0.01696s^2 - 0.05148s^3 & for \quad 0 < s < 1.9475 \\ -4.41\exp(-s/0.7176) & for \quad s > 1.9475 \end{cases}$$

Denisov, used the following universal function [18] (DP 00) to evaluate the nuclear potential

$$\phi(s) = \begin{vmatrix} 1 - \frac{s}{0.7881663} + 1.229218s^2 - 0.2234277s^3 - 0.1038769s^4 - \frac{R_1 R_2}{R_1 + R_2} \\ \times (0.1844935s^2 + 0.07570101s^3) + (I_1 + I_2)(0.04470645s^2 + 0.0334687s^3) \quad for \quad -5.65 < s < 0 \\ \left\{1 - s^2\left[0.05410106\frac{R_1 R_2}{R_1 + R_2}\exp\left(\frac{-s}{1.76058}\right) - 0.539542(I_1 + I_2)\exp\left(\frac{-s}{2.424406}\right)\right]\right\} \\ \times \exp\left(\frac{-s}{0.7881663}\right) \quad for \quad s < 0 \end{vmatrix}$$

Zhang et al., used the following universal function [19] to evaluate the nuclear potential. This universal function is called prox 13.

$$\phi(s) = P_1 \bigg/ \left[1 + \exp\left(\frac{s_0 + P_2}{P_3}\right)\right] \quad where \quad P_1 = -7.65, \; P_2 = 1.02 \; and \; P_3 = 0.89$$

The nuclear potential was studied by Qu et al., [20] using the following universal function which is also known as prox 2014.

$$\phi(s) = \begin{cases} -0.1353 + \sum_{n=0}^{5}\left(\frac{C_n}{n+1}\right)(2.5-s)^{n+1} & for \quad 0 < s < 2.5 \\ -0.0955\exp\left(\frac{2.75-s}{0.7176}\right) & for \quad s > 2.5 \end{cases}$$

Gupta et al., [21-22] used proximity potential for the preformed cluster model. Many proximity functions are available in the literature to define decay process and fusion reactions. There is no specific proximity function to study the decay process and fusion reaction in the superheavy region. There is a need to develop the new proximity function which produces experimental half-lives. In the present work, we have constructed new proximity function for superheavy nuclei region based on the experimental half-lives.

## II. Theory
### II.1 Construction of new proximity function based on the experimental alpha decay half-lives

According to the WKB theory, the probability per unit time of penetration through the barrier is $P = [1+\exp(K)]^{-1} \cong \exp(-K)$ where

$$K = \frac{2}{\hbar}\int_{r_1}^{r_2}\sqrt{2\mu(V-Q)}\,dz \tag{1}$$

The barrier penetrability is given by $P = \exp\left\{-\frac{2}{\hbar}\int_{r_1}^{r_2}\sqrt{2\mu(V-Q)}\,dz\right\}$ (2)

Here $\mu$ is the reduced mass and it is given by $\mu = mA_1A_2/A$, where $m$ is the nucleon mass and $A_1$, $A_2$ are the mass numbers of daughter and emitted clusters, respectively. For cluster/alpha decay, the turning points $r_1$ and $r_2$ are determined from the equation, $V(r_1) = V(r_2) = Q$.

After solving the above integral in eq. (1), $K$ reduces to

$$K = \frac{2\sqrt{2\mu}}{\hbar}\sqrt{(V-Q)} \times f(r_1, r_2) \tag{3}$$

Penetration probability P becomes,

$$P = \exp(-K) \quad \text{or} \quad K = -\ln P \quad \text{or} \tag{4}$$

$$\frac{2\sqrt{2\mu}}{\hbar}\sqrt{(V-Q)} \times f(r_1, r_2) = -\ln P \quad \text{or} \tag{5}$$

$$\frac{8\mu}{\hbar^2}(V-Q) \times (f(r_1, r_2))^2 = (-\ln P)^2 \quad \text{or} \tag{6}$$

$$V = \frac{(-\ln P)^2}{\frac{8\mu}{\hbar^2}(f(r_1, r_2))^2} + Q \quad \text{or} \tag{7}$$

But total potential is the sum of Coulomb and proximity potential, $V = V_N + V_C$

$$V_N + V_C = \frac{(-\ln P)^2}{\frac{8\mu}{\hbar^2}(f(r_1, r_2))^2} + Q \quad \text{or} \tag{8}$$

$$V_N = \frac{(-\ln P)^2}{\frac{8\mu}{\hbar^2}(f(r_1, r_2))^2} + Q - V_C \tag{9}$$

The above equation also represents the nuclear potential, which is based on penetration probabilities. The penetration probability intern depends on half-lives. But proximity potential ($V_N$) is composed of two parts. One depends on shape and geometry of two nuclei. The other is the universal function $\phi(s)$, only related to the distance of separation between two nuclei.

$$V_N(r) = 4\pi\gamma b \left[\frac{C_1 C_2}{C_1 + C_2}\right] \phi(s) \tag{10}$$

Comparing the equation (9) and (10), we get,

$$4\pi\gamma b \left[\frac{C_1 C_2}{C_1 + C_2}\right] \phi(s) = \frac{(-\ln P)^2}{\frac{8\mu}{\hbar^2}(f(r_1,r_2))^2} + Q - V_C \quad \text{or} \tag{11}$$

$$\phi(s) = \frac{\left[\dfrac{(-\ln P)^2}{\dfrac{8\mu}{\hbar^2}(f(r_1,r_2))^2} + Q - V_C\right]}{4\pi\gamma b \left[\dfrac{C_1 C_2}{C_1 + C_2}\right]} \tag{12}$$

The above equation gives the proximity function which depends on the penetration probability (P), energy released (Q) and Coulomb potential between emitted cluster and the daughter nucleus ($V_C$). After substituting $P = \ln 2/\upsilon T_{1/2} = 0.6931/\upsilon T_{1/2}$ in equation (12),

$$\phi(s) = \frac{\dfrac{(0.6931/\upsilon T_{1/2})^2}{8\mu/\hbar^2 (f(r_1,r_2))^2} - (Z_1 Z_2 e^2/r) + Q}{4\pi\gamma b(C_1 C_2/C_1 + C_2)} \tag{13}$$

In the above expression, $T_{1/2}$ is the alpha decay half life. We have used the available experimental half-lives of superheavy nuclei and evaluated $\phi$ for different values of separation of nuclei (s). After evaluating $\phi$ at different s, we have constructed equivalent function for $\phi$. The constructed new proximity function is given by the expression

$$\phi(s) = \begin{cases} -7.421\exp(0.534/s) & \text{for } s < 0 \\ -1.864\exp(-0.623/s) & \text{for } s > 0 \end{cases} \tag{14}$$

This proximity function is referred as "Prox. 20MS". This proximity potential is based on the experimental values of the available alpha decay half-lives in the superheavy region.

### III. Results and discussions

The nature of variation of proximity function with the separation between two nuclei is shown in figure 1. We have used the proximity functions available in the literature [13-20] and evaluated the alpha decay half-lives of superheavy nuclei whose experimental values are known. The comparison of alpha decay half-lives produced by different proximity functions along with the experimental half-lives are shown in the figure 2. The alpha decay half-lives evaluated using the present constructed proximity function is also plotted along with the corresponding experimental values in the 10[th] panel of figure 2.

To check the predictive power of constructed proximity function, we have evaluated the deviation of logarithmic half-lives with that of the experimental half-lives for different proximity functions available in the literature [13-20] and this deviation is graphically presented in figure 3. The deviation of logarithmic alpha decay half-lives produced by the present proximity function with the experimental values is also shown in the 10[th] panel of the figure 3. Again, to check the accuracy of prediction of experimental logarithmic values, we have also evaluated mean deviation of logarithmic alpha decay half-lives produced by the different proximity functions with the experiments and it is shown in figure 4. From this comparison it is observed that, the mean deviation of logarithmic half-lives are small for present constructed new proximity function. Table 1 shows the comparison of logarithmic half-lives produced by different proximity functions with that of the experiments. From this table, it is observed that present constructed new proximity function produces logarithmic half-lives close to the experiments. Hence we can conclude that, the present proximity function produces logarithmic alpha decay half-lives for superheavy nuclei.

## IV. Summary

We have constructed new proximity function particularly for interaction between two superheavy nuclei based on the experimental alpha decay half-lives. The new proximity function is used to produce the alpha decay half-lives of superheavy nuclei whose experimental values are known. The new proximity function produces the alpha decay half-lives close to the experiments. Hence we can conclude that the new proximity function can be used to study the interaction between two superheavy nuclei.

Table 1: Comparison of logarithmic half-lives produced by different proximity functions with that of the experiments

| A | Z | logT | | | | | | | | | |
|---|---|---|---|---|---|---|---|---|---|---|---|
| | | Expt. | Prox. 77 | Prox. 81 | GP77 | Bass 80 | Prox. 13 | Ng80 | Prox. 00 | Bass 77 | Prox. 14 | Prox. 20MS |
| 275 | 106 | 2.16 [23] | 2.11 | 2.11 | 2.11 | 0.43 | 2.20 | 1.40 | 3.12 | 0.21 | 3.16 | 2.11 |
| 271 | 106 | 2.06 [24] | 2.56 | 2.56 | 2.56 | 0.89 | 2.65 | 1.86 | 3.58 | 0.67 | 3.61 | 1.98 |
| 274 | 107 | 1.72 [25] | 1.60 | 1.59 | 1.60 | -0.07 | 1.68 | 0.89 | 2.62 | -0.30 | 2.66 | 1.61 |
| 275 | 108 | -0.72 [24] | 0.78 | 0.78 | 0.78 | -0.88 | 0.87 | 0.09 | 1.81 | -1.11 | 1.85 | 0.19 |
| 275 | 108 | -0.82 [23] | 0.35 | 0.35 | 0.35 | -1.31 | 0.44 | -0.34 | 1.39 | -1.54 | 1.43 | -0.24 |
| 275 | 109 | -1.01 [26] | -2.21 | -2.22 | -2.21 | -3.84 | -2.12 | -2.88 | -1.13 | -4.12 | -1.09 | -1.13 |
| 278 | 109 | 0.89 [25] | -0.08 | -0.09 | -0.08 | -1.74 | 0.00 | -0.78 | 0.95 | -1.98 | 0.99 | 0.68 |
| 279 | 110 | -0.74 [23] | -0.18 | -0.18 | -0.18 | -1.84 | -0.09 | -0.87 | 0.85 | -2.07 | 0.90 | -0.77 |
| 279 | 111 | -0.77 [26] | -1.70 | -1.70 | -1.70 | -3.34 | -1.61 | -2.37 | -0.64 | -3.59 | -0.60 | -2.31 |
| 282 | 111 | -0.29 [25] | 2.34 | 2.34 | 2.34 | 0.65 | 2.42 | 1.63 | 3.33 | 0.44 | 3.38 | 1.77 |
| 285 | 112 | 1.53 [23] | 2.14 | 2.14 | 2.14 | 0.45 | 2.23 | 1.43 | 3.13 | 0.25 | 3.18 | 1.57 |
| 283 | 112 | 0.60 [23] | 0.98 | 0.98 | 0.98 | -0.70 | 1.06 | 0.28 | 1.98 | -0.91 | 2.03 | 0.39 |
| 283 | 112 | -1.42 [24] | 1.38 | 1.38 | 1.38 | -0.30 | 1.47 | 0.68 | 2.38 | -0.51 | 2.43 | -0.80 |
| 285 | 112 | 1.53 [27] | 2.57 | 2.57 | 2.57 | 0.87 | 2.65 | 1.85 | 3.55 | 0.67 | 3.60 | 2.00 |
| 285 | 113 | 0.74 [25] | 0.69 | 0.68 | 0.69 | -0.99 | 0.77 | -0.01 | 1.69 | -1.20 | 1.74 | 0.71 |
| 286 | 113 | -1.70 [25] | 1.02 | 1.02 | 1.02 | -0.66 | 1.11 | 0.32 | 2.02 | -0.87 | 2.07 | -0.44 |
| 283 | 113 | -1.00 [26] | -0.37 | -0.37 | -0.37 | -2.03 | -0.28 | -1.06 | 0.66 | -2.25 | 0.70 | -0.97 |
| 288 | 114 | -1.74 [28] | 1.14 | 1.14 | 1.14 | -0.55 | 1.22 | 0.43 | 2.13 | -0.76 | 2.18 | -0.56 |
| 287 | 114 | -3.32 [24] | 0.60 | 0.60 | 0.60 | -1.08 | 0.69 | -0.10 | 1.60 | -1.29 | 1.66 | 0.02 |
| 289 | 114 | -1.28 [27] | 1.15 | 1.15 | 1.15 | -0.54 | 1.23 | 0.44 | 2.13 | -0.75 | 2.19 | -0.57 |
| 288 | 114 | -0.20 [27] | 0.79 | 0.79 | 0.79 | -0.89 | 0.88 | 0.09 | 1.79 | -1.10 | 1.84 | 0.21 |
| 286 | 114 | -0.54 [27] | 0.59 | 0.59 | 0.59 | -1.09 | 0.68 | -0.11 | 1.60 | -1.29 | 1.65 | 0.50 |
| 288 | 114 | -0.20 [27] | 0.79 | 0.79 | 0.79 | -0.89 | 0.88 | 0.09 | 1.79 | -1.10 | 1.84 | -0.21 |
| 286 | 114 | -0.54 [27] | 0.59 | 0.59 | 0.59 | -1.09 | 0.68 | -0.11 | 1.60 | -1.29 | 1.65 | 0.01 |
| 289 | 114 | 0.43 [23] | 0.74 | 0.74 | 0.74 | -0.95 | 0.83 | 0.03 | 1.73 | -1.16 | 1.79 | 0.16 |
| 288 | 114 | -0.10 [23] | 0.39 | 0.39 | 0.39 | -1.29 | 0.47 | -0.31 | 1.39 | -1.50 | 1.44 | -0.20 |
| 287 | 114 | -0.29 [23] | 0.19 | 0.19 | 0.19 | -1.48 | 0.28 | -0.51 | 1.20 | -1.70 | 1.25 | -0.40 |
| 286 | 114 | -0.80 [23] | -0.32 | -0.32 | -0.32 | -1.99 | -0.23 | -1.01 | 0.70 | -2.20 | 0.75 | -0.91 |
| 289 | 114 | 0.43 [23] | 0.74 | 0.74 | 0.74 | -0.95 | 0.83 | 0.03 | 1.73 | -1.16 | 1.79 | 0.16 |
| 288 | 114 | -0.10 [23] | 0.39 | 0.39 | 0.39 | -1.29 | 0.47 | -0.31 | 1.39 | -1.50 | 1.44 | -0.20 |
| 289 | 115 | -0.66 [25] | -0.30 | -0.30 | -0.30 | -1.97 | -0.21 | -0.99 | 0.71 | -2.18 | 0.77 | -0.89 |
| 287 | 115 | -1.49 [26] | -1.06 | -1.07 | -1.06 | -2.72 | -0.97 | -1.74 | -0.04 | -2.94 | 0.02 | -1.66 |

| 290 | 115 | -1.80 [25] | 0.70  | 0.70  | 0.70  | -0.99 | 0.79  | -0.01 | 1.69  | -1.19 | 1.75  | 0.12  |
| 290 | 116 | -1.82 [29] | -1.46 | -1.46 | -1.46 | -3.11 | -1.37 | -2.14 | -0.43 | -3.34 | -0.37 | -2.06 |
| 291 | 116 | -1.20 [29] | -1.17 | -1.17 | -1.17 | -2.83 | -1.08 | -1.86 | -0.15 | -3.06 | -0.09 | -1.77 |
| 292 | 116 | -1.74 [30] | -0.97 | -0.97 | -0.97 | -2.64 | -0.88 | -1.66 | 0.04  | -2.86 | 0.10  | -1.56 |
| 293 | 116 | -0.28 [29] | -0.63 | -0.63 | -0.63 | -2.30 | -0.54 | -1.33 | 0.37  | -2.52 | 0.44  | -1.22 |
| 286 | 116 | -0.89 [24] | 0.83  | 0.83  | 0.83  | -0.84 | 0.92  | 0.14  | 1.84  | -1.04 | 1.89  | -0.25 |
| 294 | 117 | -0.75 [25] | -1.07 | -1.07 | -1.07 | -2.74 | -0.98 | -1.76 | -0.07 | -2.96 | 0.00  | -1.66 |
| 293 | 117 | -1.85 [25] | -1.61 | -1.62 | -1.61 | -3.27 | -1.53 | -2.30 | -0.60 | -3.50 | -0.53 | -2.21 |
| 294 | 118 | -3.05 [24] | -2.46 | -2.47 | -2.46 | -4.11 | -2.37 | -3.14 | -1.43 | -4.35 | -1.37 | -3.07 |
| 294 | 118 | -2.74 [23] | -2.85 | -2.85 | -2.85 | -4.49 | -2.76 | -3.52 | -1.81 | -4.73 | -1.75 | -3.46 |

Fig.1: Variation of φ with that of r

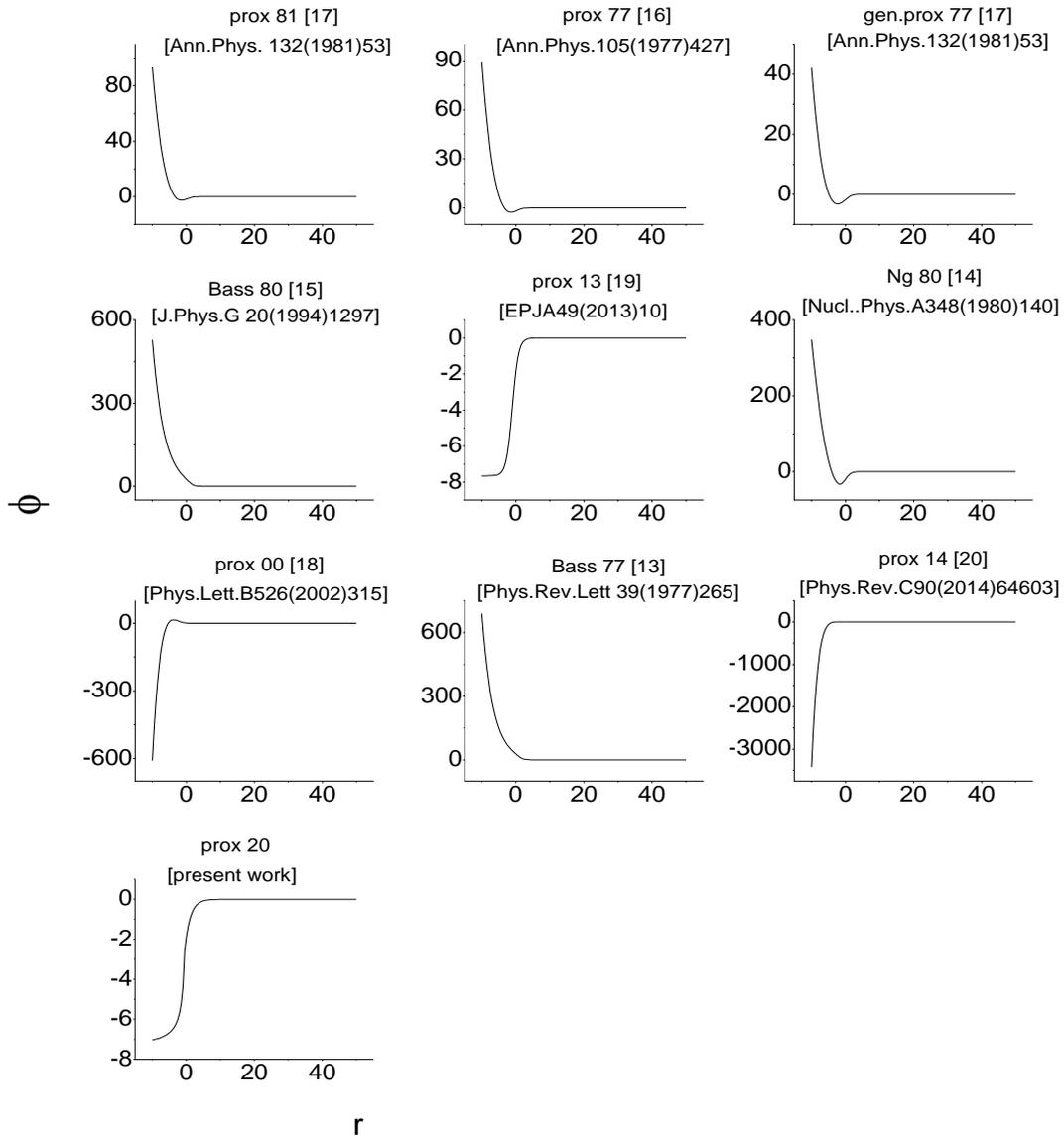

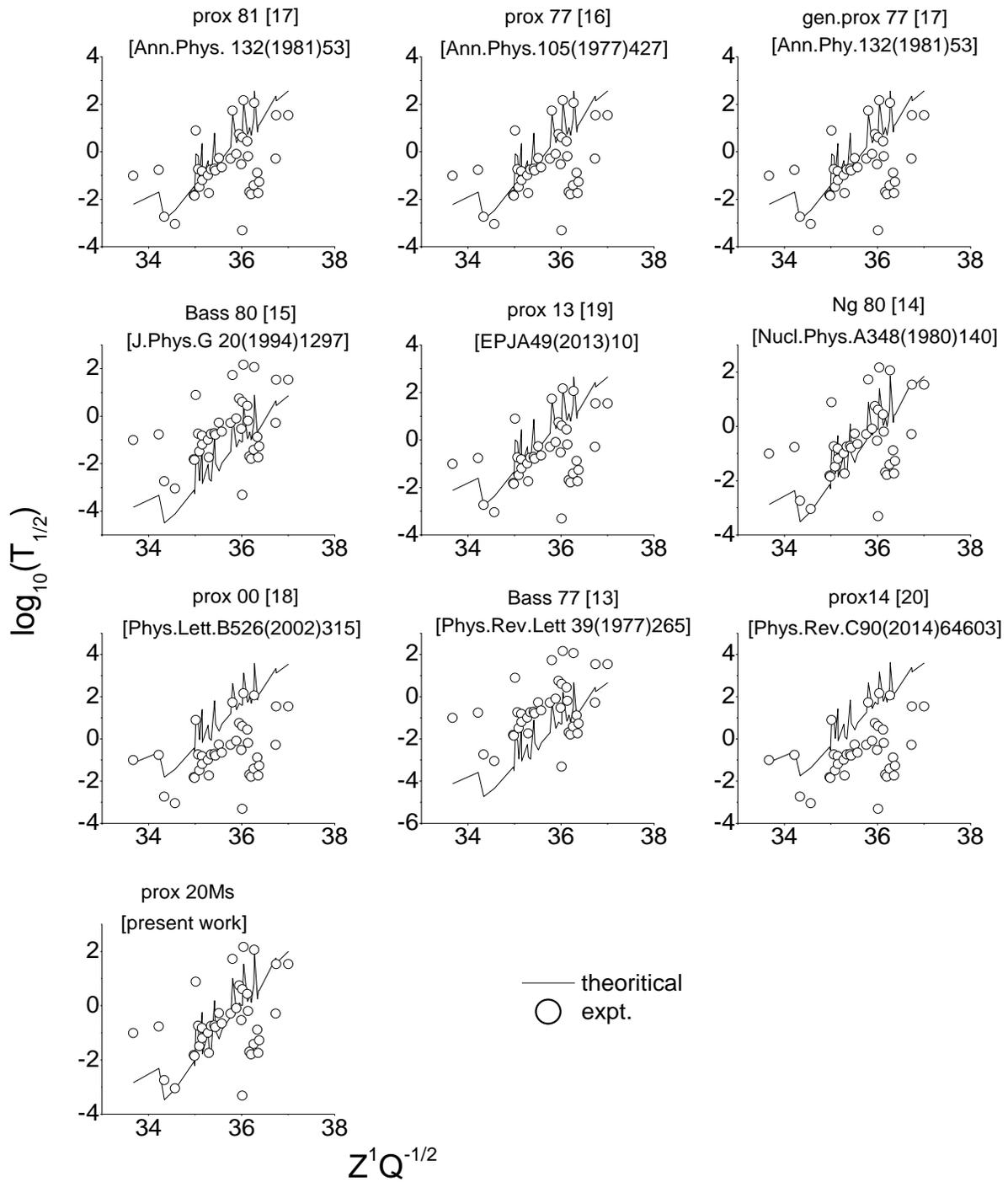

Fig.2: Variation of $\log T_{1/2}$ as a function of $ZQ^{-1/2}$

Fig.3: Variation of deviation ($\log T_{1/2} - \log T_{1/2}.expt$) with reference to the experimental value as a function of $ZQ^{-1/2}$

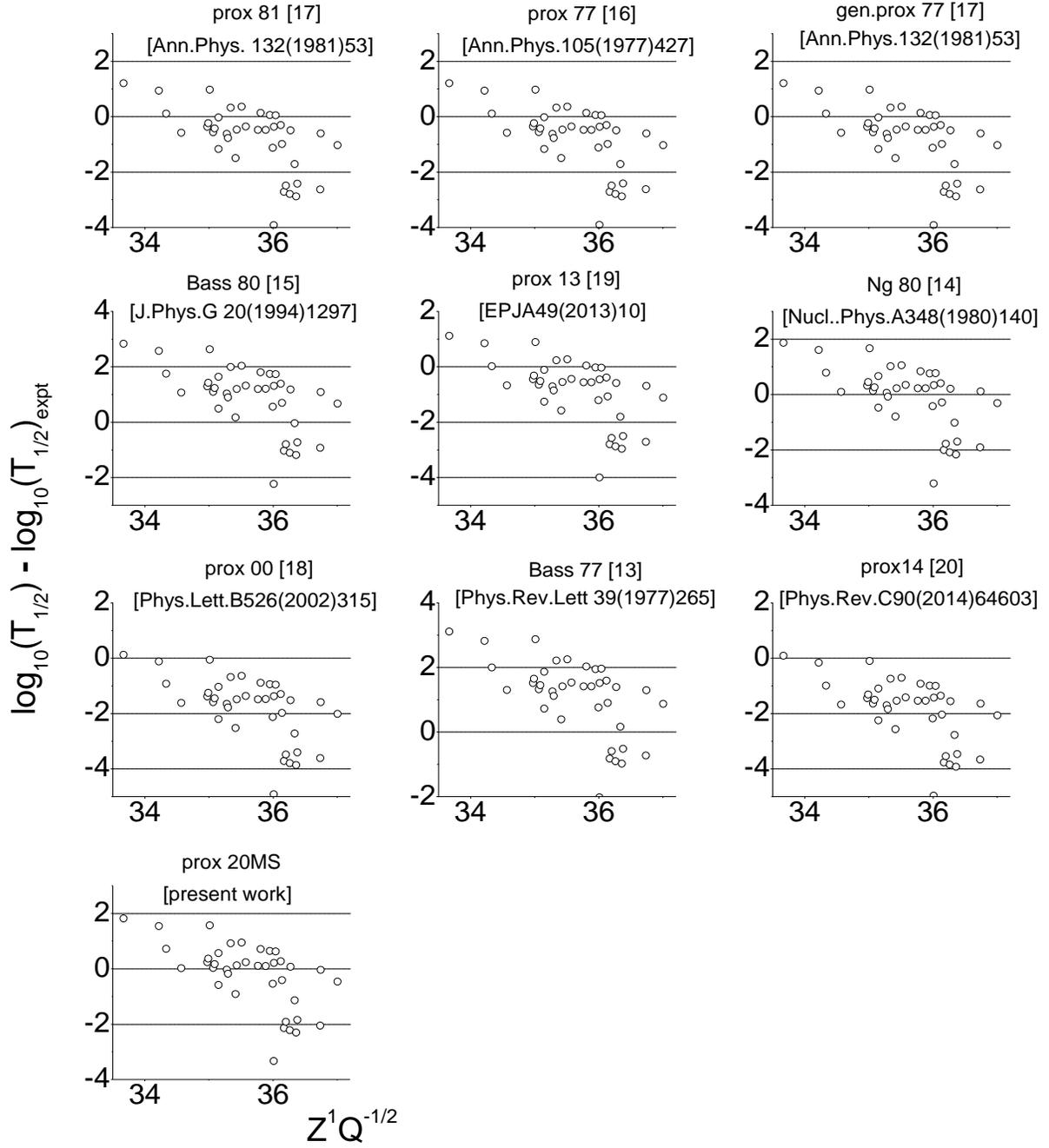

Fig.4: Comparison of mean deviation of log$T_{1/2}$ with that of experiments for different proximity functions available in the literature.

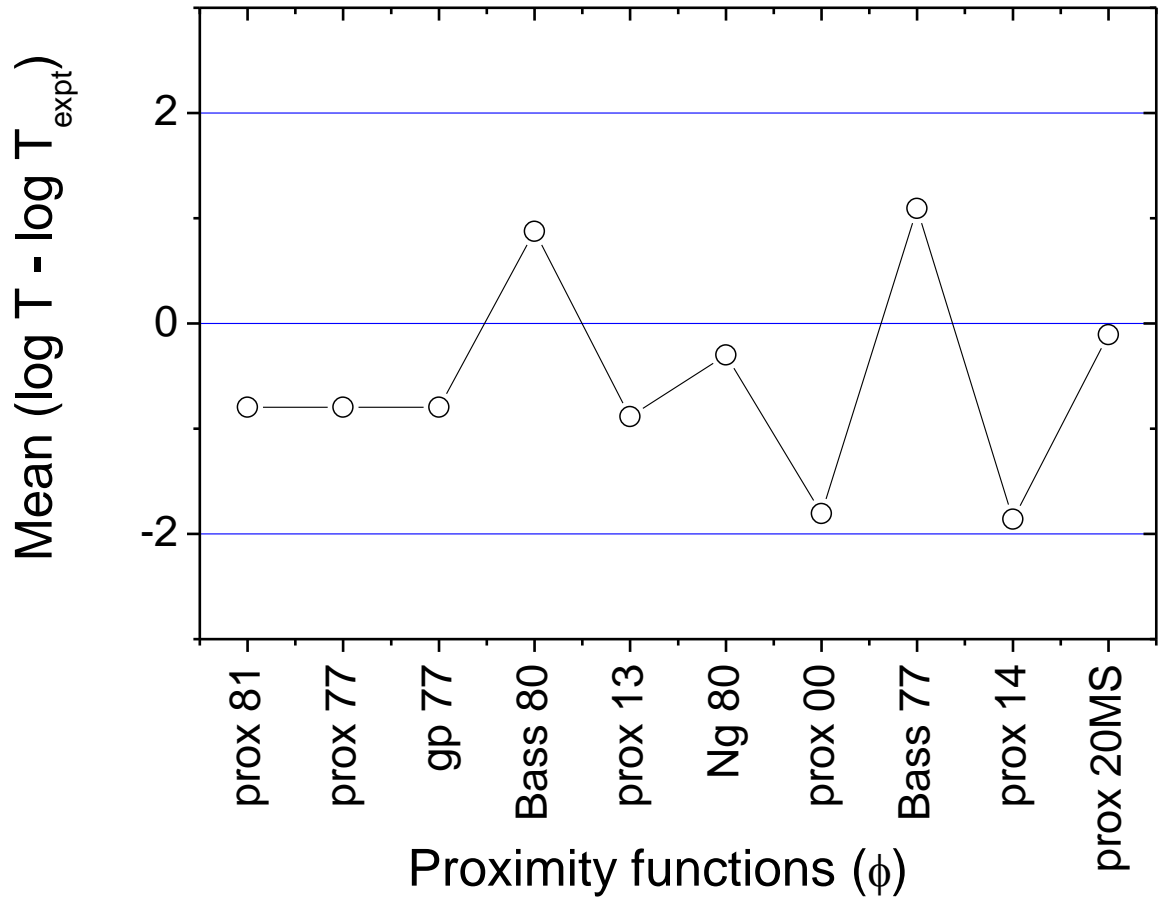